\documentclass[prc,aps,superscriptaddress]{revtex4}  
\usepackage{url,hyperref,lineno,microtype}
\usepackage{amsfonts}
\usepackage{amssymb}
\usepackage{amsmath}
\usepackage{graphicx}

\begin{document}

\title{Non-Symmetrized Hyperspherical Harmonics Method for Non-Equal Mass
       Three-Body Systems} 

\author{A. Nannini}
\affiliation{Dipartimento di Fisica ``Enrico Fermi'', Universit\`a
di Pisa, Largo Bruno Pontecorvo 3 - I-56127 Pisa, Italy}
\affiliation{Istituto Nazionale di Fisica Nucleare, Sezione di Pisa,\\
Largo Bruno Pontecorvo 3 - I-56127 Pisa, Italy}
\author{L.E. Marcucci}
\affiliation{Dipartimento di Fisica ``Enrico Fermi'', Universit\`a
di Pisa, Largo Bruno Pontecorvo 3 - I-56127 Pisa, Italy}
\affiliation{Istituto Nazionale di Fisica Nucleare, Sezione di Pisa,\\
Largo Bruno Pontecorvo 3 - I-56127 Pisa, Italy}

\begin{abstract}
  The non-symmetrized hyperspherical harmonics method for a three-body system,
  composed by two particles having equal masses, but different from the mass
  of the third particle, is reviewed and applied to
  the $^3$H, $^3$He nuclei and $^3_{\Lambda}$H hyper-nucleus,
  seen respectively as $nnp$, $ppn$ and $NN\Lambda$ three-body
  systems.
  The convergence of the method is first
  tested in order to estimate its accuracy.
  Then, the difference of binding energy between $^3$H and $^3$He due to the
  difference of the proton and the neutron masses
  is studied using several central spin-independent and spin-dependent
  potentials. Finally, the $^3_{\Lambda}$H hypernucleus binding energy
  is calculated using different $NN$ and $\Lambda N$ potential models.
  The results have been compared with those present in the literature,
  finding a very nice agreement.
\end{abstract}

\maketitle

\section{Introduction}
\label{sec:intro}

The hyperspherical harmonics (HH) method has been widely applied in the
study of the bound states of few-body systems, starting from $A=3$
nuclei~\cite{kie2008,lei2013}. Usually, the use of the HH basis is preceded
by a symmetrization procedure that takes into account the fact that protons and
neutrons are fermions, and the wave function
has to be antisymmetric under exchange of any pair of these particles. 
For instance, for $A=3$, antisymmetry is guaranteed by writing the wave function
as 
\begin{equation}
  \Psi=\sum_p \Psi_p \ ,
  \label{eq:psi}
\end{equation}
$p=1,2,3$ corresponding to the three
different particle permutations~\cite{kie2008}.
However, it was shown in Refs.~\cite{gat2009a,gat2009b,gat2011,def2013,def2014} that
this preliminary step is in fact not strictly necessary, since,
after the diagonalization of the Hamiltonian, the eigenvectors turn out
to have a well-defined symmetry
under particle permutation.
In this second version, the method is known as non-symmetrized hyperspherical
harmonics (NSHH) method.
    As we will also show below, the prize to pay for the
    non-antisymmetrization is that a quite larger number of the expansion
    elements are necessary with respect to the ``standard'' HH method.
However, the NSHH method has the advantage to reduce the computational effort
due to the symmetrization procedure, and, moreover, the same expansion can be
easily re-arranged for systems of different particles with different
masses.
In fact, the steps to be done within the NSHH method from the case of
equal-mass to the case of non-equal mass particles are quite straightforward
and will be illustrated below. In this work, we apply the NSHH method to study 
the $^3$H,
$^3$He and $^3_{\Lambda}$H systems, seen as $nnp$, $ppn$, $NN\Lambda$
respectively (we used the standard notation of $N$ for nucleon
and $Y$ for hyperon).
In order to test our method,  we study the first two systems listed above with
five different potential models, and the hypernucleus with three
potential models. We start with simple central spin-independent
$NN$ and $YN$ interactions, and then we move to 
central spin-dependent potentials.
To be noticed that none of the interactions considered is
realistic. Furthermore, we do not include three-body forces.
Therefore, the comparison of our results with the experimental data
is meaningless. However, the considered interactions are useful to test step
by step our method and to compare with results obtained in the literature.

The paper is organized as follows: in Section~\ref{sec:form} we describe the
NSHH method, in Section~\ref{sec:res} we discuss the results obtained for the
considered nuclear systems. Some concluding remarks and an outlook
are presented in Section~\ref{sec:conc}.
\section{Theoretical formalism}
\label{sec:form}

We briefly review the formalism of the present calculation. We start
by introducing the Jacobi coordinates for a system of $A=3$ particles, with mass
$m_{i}$, position ${\bf r}_{i}$, and momentum ${\bf p}_{i}$.
By defining ${\bf x}_i=\sqrt{m_i} \ {\bf r}_i$~\cite{ros},
they are taken as a linear combination of ${\bf x}_i$, i.e. 
\begin{equation}
{\bf y}_{i}=\sum\limits_{j=1}^A c_{ij}{\bf x}_{j} \ ,
\label{eq:yij}
\end{equation}
where the coefficients $c_{ij}$ need to satisfy the following conditions~\cite{ros}
\begin{eqnarray}
\sum\limits_{i=1}^3 \ c_{ji}c_{ji}&=&\frac{1}{M}  \ \ \ \ (j=1,2) \ ,
\label{eq:cond1} \\
\sum\limits_{i=1}^3 c_{ji}c_{ki}&=&0 \ \ \ \  (j\neq{k}=1,2) \ .
\label{eq:cond2}
\end{eqnarray}
Here $M$ is a reference mass.
The advantage of using Eqs.~(\ref{eq:yij})--(\ref{eq:cond2}) is that
the kinetic energy operator can be cast in the form
\begin{equation}
  T=-\frac{\hbar^2}{2m_{tot}}\nabla^2_{\textbf{y}_{3}}-\frac{\hbar^2}{2M}
  (\nabla^2_{\textbf{y}_{1}} + \nabla^2_{\textbf{y}_{2}}) \ ,
\label{eq:T}
\end{equation}
where $\textbf{y}_{3}$ is the center-of-mass coordinate.
For a three-body system, there are three possible
permutations of the particles. Therefore, the Jacobi coordinates
depend on this permutations. For $p=3$, i.e.\ $i,j,k=1,2,3$, the Jacobi coordinates
are explicitly given by
\begin{eqnarray}\label{coordinates}
\textbf{y}_2^{(3)}&=&-\sqrt{\frac{m_{2}}{M(m_{1}+m_{2})}}\textbf{x}_{1}+\sqrt{\frac{m_{1}}{M(m_{1}+m_{2})}}\textbf{x}_{2} \ ,
\nonumber\\
\textbf{y}_1^{(3)}&=&-\sqrt{\frac{m_{1}m_{3}}{Mm_{tot}(m_{1}+m_{2})}}\textbf{x}_{1}-\sqrt{\frac{m_{2}m_{3}}{Mm_{tot}(m_{1}+m_{2})}}\textbf{x}_{2}+\sqrt{\frac{m_{1}+m_{2}}{Mm_{tot}}}\textbf{x}_{3} \ . \ \ \ \
\label{eq:y12}
\end{eqnarray}
They reduce to the familiar expressions for equal-mass particles
when $m_1=m_2=m_3=M$
(see for instance Ref.~\cite{kie2008}).
We then introduce the hyperspherical coordinates, by replacing,
in a standard way, the moduli of $\textbf{y}_{1,2}^{(3)}$
by the hyperradius and one hyperangle, given by
\begin{eqnarray}
 && \rho^2={\textbf{y}_1^{(p)}}^2+{\textbf{y}_2^{(p)}}^2
  \label{eq:rho} \ , \\
 &&\tan\phi^{(p)}=\frac{{y}_1^{(p)}}{{y}_2^{(p)}}
  \label{eq:phi} \ .
\end{eqnarray}
To be noticed that the hyperangle $\phi^{(p)}$ depends on the permutation
$p$, while the hyperradius $\rho$ does not. The well-known
advantage of using the hyperspherical coordinates is that
the Laplace operator can be cast in the form~\cite{ros}
\begin{equation}
  \label{eq:nabla}
  \nabla^2  =  \nabla^2_{\textbf{y}_{1}}+\nabla^2_{\textbf{y}_{2}}= 
  \frac{\partial^2}{\partial \rho^2 }+\frac{5}{\rho}\frac{\partial}{\partial \rho}
  + \frac{\Lambda^2(\Omega^{(p)})}{\rho^2} \ ,
\end{equation}
where ${\Lambda^2(\Omega^{(p)})}$ is called the grand-angular momentum operator,
and is explicitly written as
\begin{eqnarray}
  \label{eq:grandangolare}
\Lambda^2(\Omega^{(p)})&=&\frac{\partial^2}{\partial \phi^{(p)2}}-\frac{\hat{\ell}^2_{1}(\hat{\textbf{y}}_{1}^{(p)})}{\sin^2 \phi^{(p)}}-\frac{\hat{\ell}^2_{2}(\hat{\textbf{y}}_{2}^{(p)})}{\cos^2 \phi^{(p)}}+2\bigg[\cot\phi^{(p)}-\tan\phi^{(p)}\bigg]\frac{\partial}{\partial \phi^{(p)}} \ .
\end{eqnarray}
Here $\hat{\ell}^2_{1}$ and $\hat{\ell}^2_{2}$ are the (ordinary) angular momentum operators
associated with the Jacobi vectors ${\textbf{y}}_{1}^{(p)}$ and ${\textbf{y}}_{2}^{(p)}$ respectively, and
$\Omega^{(p)}\equiv (\hat{\textbf{y}}_{1}^{(p)},\hat{\textbf{y}}_{2}^{(p)},\phi^{(p)})$.
The HH functions are the eigenfunctions of the grand-angular momentum operator
$\Lambda^2(\Omega^{(p)})$, with eigenvalue $-G(G+4)$, i.e.\
\begin{equation}
  \label{eq:autovalorieq}
\Lambda^2(\Omega^{(p)})Y_{G}(\Omega^{(p)})=-G(G+4)Y_{G}(\Omega^{(p)}) \ .
\end{equation}
Here the HH function $Y_{G}(\Omega^{(p)})$ is defined as
\begin{eqnarray}
  Y_{G}(\Omega^{(p)})&=&N_{n}^{\ell_{1},\ell_{2}}(\cos\phi^{(p)})^{\ell_{2}}
  (\sin\phi^{(p)})^{\ell_{1}}Y_{\ell_{1}m_{1}}(\hat{\textbf{y}}_{1}^{(p)})
  Y_{\ell_{2}m_{2}}(\hat{\textbf{y}}_{2}^{(p)})  \nonumber \\ 
  &\times& \ P_{n}^{\ell_{1}+\frac{1}{2},\ell_{2}+\frac{1}{2}}(\cos 2\phi^{(p)}) \ ,
  \label{eq:yg}
\end{eqnarray}
with $N_{n}^{\ell_{1},\ell_{2}}$ a normalization factor~\cite{ros} and
\begin{equation}
G=2n + \ell_{1} + \ell_{2} \ , \ \ \  \ n=0,1,\dots \ ,
\label{eq:G}
\end{equation}
is the so-called grand-angular momentum.
We remark that the HH functions depend on the considered permutation
via $\Omega^{(p)}$.
It is useful to combine the HH functions in order to assign them
a well defined total orbital angular momentum $\Lambda$.
Using the Clebsch-Gordan coefficients, we introduce the functions
$H_{G}(\Omega^{(p)})$ as
\begin{eqnarray}
  H_{G}(\Omega^{(p)})&=& \sum_{m_{1},m_{2}}Y_{G}(\Omega^{(p)})(\ell_{1}m_{1}\ell_{2}m_{2}|
  \Lambda\Lambda_{z})\nonumber \\ 
  &\equiv&[Y_{\ell_{1}}(\hat{\textbf{y}}_1^{(p)})
    Y_{\ell_{2}}(\hat{\textbf{y}}_2^{(p)})]_{\Lambda,\Lambda_{z}} 
  P_{n}^{\ell_1,\ell_2}(\phi^{(p)}) \ ,
\label{eq:HG}
\end{eqnarray}
where $[G]$ stands for $ [\ell_1,\ell_2,\Lambda,n]$, and
\begin{equation}
  P_{n}^{\ell_1,\ell_2}(\phi^{(p)}) = N_{n}^{\ell_1,\ell_2} (\cos\phi^{(p)})^{\ell_2}
  (\sin\phi^{(p)})^{\ell_1}P_n^{\ell_1+\frac{1}{2},\ell_2+\frac{1}{2}}(\cos{2\phi^{(p)}})
  \ .
  \label{eq:pn2}
\end{equation}
We now consider our system made of three particles, two with equal masses,
different from the mass of the third particle. We choose to fix
the two equal mass particles in position 1 and 2, and we set
the third particle with different mass as particle 3.
Therefore, we will work with the Jacobi and hyperspherical
coordinates with fixed permutation $p=3$.

The wave function that describes our system can now be cast in the form
\begin{equation}
\Psi= \sum_{\{G\}} 
BH^J_{\{G\}}(\Omega^{(3)}) \ 
u_{\{G\}}(\rho) \ ,
\label{eq:expansion}
\end{equation}
where $u_{\{G\}}(\rho)$ is a function of only the hyperradius $\rho$, and 
$BH^J_{\{G\}}(\Omega^{(3)})$ is given by Eq.~(\ref{eq:HG})
multiplied by the spin part, i.e. 
\begin{equation}\label{spin1}
BH^J_{\{G\}}(\Omega^{(3)}) =\sum_{\Lambda_z, \Sigma_z} H_{[G]}(\Omega^{(3)}) \times
\left[\left[\frac{1}{2}\otimes\frac{1}{2}\right]_{S,s} \otimes \frac{1}{2} \right]_{\Sigma,\Sigma_z} \times
(\Lambda\Lambda_z,\Sigma\Sigma_z|J J_z)\ .
\end{equation}
Here 
$S$ is the spin of the first couple with third component $s$, 
$\Sigma$ is the total spin of the system and $\Sigma_z$ its third component,
and $\{G\}$  now stands for $\{l_1,l_2,n,\Lambda,S,\Sigma\}$.
To be noticed that the $ {L}{S}$-coupling scheme is used, so that
the total spin of the system is combined, using the Clebsh-Gordan
coefficient $(\Lambda\Lambda_z,\Sigma\Sigma_z|J J_z)$,
with the total orbital angular momentum to give the total spin
$J$. Furthermore,
(i)
$\ell_1,\ell_2$ and $n$ are taken such that Eq.~(\ref{eq:G}) is satisfied for
$G$ that runs from $G^{min}=\ell_1+\ell_2$ to a given $G^{max}$,
to be chosen in order to reach the desired accuracy, and
(ii) we have imposed $\ell_1+\ell_2=$ even, since the systems
under consideration have positive parity. The possible values for
$\Lambda, \Sigma,$ and $G^{min}$, which together with $G^{max}$ identify a channel,
are listed in
Table~\ref{tab:channel} for a system with $J^\pi=1/2^+$.
Note that, since we are using central
potentials, only the first channel
of Table~\ref{tab:channel} will be in fact necessary.
\begin{table}[t] \centering
	\begin{tabular}{ccccc}
		\hline 
		&$ch$&$\Lambda$&$\Sigma$&$G^{min}$\\ \hline
		&$1$&$0$&$1/2$&0 \\
		&$2$&$1$&$1/2$&2 \\
		&$3$&$1$&$3/2$&2 \\
		&$4$&$2$&$3/2$&2 \\
		\hline 
	\end{tabular}
	\caption{List of the channels for a $J^{\pi}=1/2^+$ system.
          $\Lambda$ and $\Sigma$ are the total orbital angular
          momentum and the total spin of the nuclei. See text for more details.}
	\label{tab:channel}
\end{table}
%
%

In the present work, the hyperradial function is itself expanded on a suitable
basis, i.e. a set of generalized Laguerre polynomials~\cite{kie2008}.
Therefore, we can write
\begin{equation}
u_{\{G\}}(\rho)= \sum_{l}c_{\{G\},l}\,f_l(\rho) \ ,   
\label{eq:uf}
\end{equation}
where $c_{\{G\},l}$ are unknown coefficients, and
\begin{equation}
  \label{eq:laguerre}
f_{l}(\rho)= \sqrt{\frac{l!}{(l+5)!}}\gamma^{3} \
{^{(5)} L_{l}(\gamma \rho)}e^{-\frac{\gamma}{2}\rho} \ .
\end{equation}
Here $^{(5)}L_l(\gamma\rho)$ are generalized Laguerre polynomials,
and the numerical factor in front of them is chosen so that
$f_{l}(\rho)$ are normalized to unit. Furthermore, $\gamma$ is a non-linear
parameter, whose typical values are in the range $(2-5)$ fm$^{-1}$.
The results have to be stable against $\gamma$, as we will show
in Section~\ref{sec:res}.
With these assumptions, the functions $ f_{l}(\rho)$ go to zero for
$\rho\rightarrow\infty$, and constitute an orthonormal basis.

By using Eq.~(\ref{eq:uf}), the wave function can now be cast in the form
\begin{equation}
\Psi= \sum_{\{G\}} \sum_{l=1}^{N_{max}}
\,c_{\{G\},l}\, BH^J_{\{G\}}(\Omega) \,
f_{l}(\rho) \ ,
\label{eq:wf}
\end{equation}
where we have dropped the superscript $(3)$ in $\Omega^{(3)}$ to simplify
the notation, and we have indicated with $N_{max}$ the maximum number of
Laguerre polynomials in Eq.~(\ref{eq:uf}).

In an even more compact notation, we can write
\begin{equation}
\Psi=\sum_{\xi} c_{\xi}\Psi_{\xi} \ ,
\label{eq:psi_compact}
\end{equation}
where $\Psi_{\xi}$ is a complete set of states, and $\xi$ is the index
that labels all the quantum numbers defining the basis elements.
The expansion coefficients $c_{\xi}$ can be determined using the
Rayleigh-Ritz variational principle~\cite{kie2008}, which states that
\begin{equation}
\langle
\delta_c \Psi | H-E | \Psi
\rangle =0 \ ,
\label{eq:rr-var}
\end{equation}
where $\delta_c \Psi$ denotes the variation of the wave function
with respect to the coefficients $c_{\xi}$. 
By doing the differentiation, the problem is then reduced to a generalized
eigenvalue-eigenvector problem of the form
\begin{equation}
\sum_{\xi'}	\langle
\Psi_{\xi}|H-E|\Psi_{\xi'}
\rangle c_{\xi'}=0 \ ,
\label{eq:equations-lanczos}
\end{equation}
that is solved using the Lanczos diagonalization algorithm~\cite{che1986}.
The use of the Lanczos algorithm is dictated by the large size
($\sim 50000\times 50000$) of the involved matrices (see below).

All the computational problem is now shifted in having to calculate
the norm, kinetic energy and potential energy matrix elements. One of the
advantage of using a fixed permutation is that the norm and kinetic
energy matrix elements are or analytical, or involve
just a one-dimensional integration. In fact, they are written as
\begin{eqnarray}
N_{\{G'\},k;\{G\},l} &\equiv& \langle\Psi_{\xi'} | \Psi_{\xi}
\rangle=J \ \delta_{\xi,\xi'}
\ , \label{eq:norma} \\
T_{\{G'\},k;\{G\},l} &\equiv& \langle\Psi_{\xi'} |T |\Psi_{\xi}
\rangle=
-\frac{\hbar^2}{2M} \ J \ \delta_{\{G\}, \{G'\}} \ \int d\rho \ \rho^{5}
f_k(\rho) \\ \nonumber
&\times& \left[
-G(G+4)\frac{f_l(\rho)}{\rho^2} + 5 \ \frac{f'_l(\rho)}{\rho} + f''_l(\rho)
\right] \ ,
\label{eq:cinetica}
\end{eqnarray}
where $J$ is the total Jacobian of the transformation, given by
\begin{equation}
J=\bigg({M \sqrt{\frac{m_{tot}}{m_1 m_2 m_3}}}\bigg)^3 \ ,
\end{equation}
and ${f'_l(\rho)}$ and ${f''_l(\rho)}$ are, respectively,
the first and the second
derivatives of the functions ${f_l(\rho)}$ defined in
Eq.~(\ref{eq:laguerre}).

The potential matrix elements in Eq.~(\ref{eq:equations-lanczos})
can be written as
\begin{equation}
  V_{\{G'\},k;\{G\},l}\equiv \langle\Psi_{\xi'} |V_{12}+V_{23}+V_{13} |
  \Psi_{\xi} \rangle \ ,
  \label{eq:potential}
\end{equation}
with $V_{ij}$ indicating the two-body interaction between particle $i$
and particle $j$. Note that in the present work we do not consider
three-body forces.
Since it is easier to evaluate the matrix elements of $V_{ij}$
when the Jacobi coordinate $\textbf{y}_2$ is proportional to
$\textbf{r}_i-\textbf{r}_j$, we proceed as follows. We make use of the
fact that the hyperradius is permutation-independent, and we use the fact that
the HH function written in terms of $\Omega^{(p)}$ can
be expressed as function of the HH written using $\Omega^{(p')}$,
with $p'\ne p$. Basically it can be shown that~\cite{kie2008}
\begin{equation}
H_{[G]}(\Omega^{(p)})=\sum_{[G']} {a_{[G],[G']}^{(p \rightarrow p'),G,\Lambda}}
{H_{[G']}}(\Omega^{(p')}) \ ,
\label{eq:HaH}
\end{equation}
where the grandangular  momentum $G$
and the total angular momentum $\Lambda$ remain constant,
i.e.\ $G=G'$ and $\Lambda=\Lambda'$, but we have
$[G] \ne [G']$, since all possible combinations of
$\ell_1,\ell_2,n$ are allowed. The spin-part written in terms of
permutation $p$ can be easily expressed in terms of permutation $p'$
via the standard $6j$ Wigner coefficients~\cite{Edmonds}.
The transformation coefficients $a_{[G],[G']}^{(p \rightarrow p'),G}$
can be calculated, for $A=3$, through the Raynal-Revai
recurrence relations~\cite{ray1970}. Alternately  we can use the
orthonormality of the HH basis~\cite{kie2008}, i.e.\
\begin{equation}
  a_{[G],[G']}^{(p \rightarrow p'),G,\Lambda}= \int d \Omega^{(p')} \
  [H_{[G']}(\Omega^{(p')})]^\dagger \  H_{[G]}(\Omega^{(p)}) \ . 	
\label{eq:a}
\end{equation}
Their explicit expression can be found for instance
in Ref.~\cite{kie2008} as is reported in the Appendix
for completeness.
The final expression for the potential matrix elements
is given by
\begin{eqnarray}
\langle\Psi_{\xi'} |V_{12}+V_{23}+V_{13} |\Psi_{\xi} \rangle
&=&	J  \int d\rho \ \rho^5 f_{k}(\rho)
f_{l}(\rho) \nonumber \\
&\times&
\bigg\{
\int d \Omega^{(3)}  BH_{\xi'}^{\dagger}(\Omega^{(3)}) V_{12} BH_{\xi}(\Omega^{(3)}) \nonumber \\
&+&  \sum_{\xi''}\sum_{\xi'''}  \bigg[
a_{\xi'\rightarrow\xi'''}^{(3 \rightarrow 1),G',\Lambda'} 
a_{\xi\rightarrow\xi''}^{(3 \rightarrow 1),G,\Lambda}
\nonumber\\ 
&\times&
\int d \Omega^{(1)} BH_{\xi'''}^{\dagger}(\Omega^{(1)})  V_{23} 
BH_{\xi''}(\Omega^{(1)}) \nonumber \\
&+& a_{\xi'\rightarrow\xi'''}^{(3 \rightarrow 2),G',\Lambda'} 
a_{\xi\rightarrow\xi''}^{(3 \rightarrow 2),G,\Lambda}
\nonumber\\
&\times&
\int d \Omega^{(2)}
BH_{\xi'''}^{\dagger}(\Omega^{(2)})  V_{13} 
BH_{\xi''}(\Omega^{(2)}) \bigg] \bigg\} \ .
\label{eq:potential-2}
\end{eqnarray}
It is then clear the advantage of using the NSHH method also for the
calculation of the potential matrix elements, as in fact all what is
needed is the calculation of one integral of the type
\begin{equation}
  I(\rho)= \int d\rho \ \rho^5 f_{k}(\rho) f_{l}(\rho)
  \int d \Omega^{(p)}
BH_{\xi'''}^{\dagger}(\Omega^{(p)})  V_{ij} 
BH_{\xi''}(\Omega^{(p)}) \ ,
\label{eq:i_ijk}
\end{equation}
with $p$ the permutation corresponding to the order $i,j,k$.

\section{Results}
\label{sec:res}

We present in this section the results obtained with the
NSHH method described above.
In particular, we present in Section~\ref{subsec:conv}
the study of the convergence of the method, in the case of the
triton binding energy, calculated with $m_p=m_n$. We then present
in Section~\ref{subsec:3h-3he} the results for the triton and $^3$He
binding energy, when $m_p\ne m_n$. In Section~\ref{subsec:hyper3h}
we present the results of the hypertriton.

The potential models used in our study are central spin-independent
and spin-dependent. In particular, 
the $^3$H and $^3$He systems have been investigated using the
spin-independent Volkov~\cite{vol1965},  Afnan-Tang~\cite{afn1968} and
Malfliet-Tjon~\cite{mal1969} potential models, and the two spin-dependent
Minnesota~\cite{tho1977} and Argonne AV4$'$~\cite{wir2002}
potential models. Note that the AV4$'$ potential is a reprojection of the
much more realistic Argonne AV18~\cite{wir1995} potential model.
In the case of the hypernucleus $_\Lambda^3$H, we have used the Gaussian
spin-independent central potential of Ref.~\cite{cla1985}, and
two spin-dependent potentials: the first one, labeled MN9~\cite{phd2017},
combines a Minnesota~\cite{tho1977} potential for the $NN$ interaction with
the $S=1$ component of the same Minnesota potential multiplied by a factor
0.9 for the $\Lambda N$ interaction. The second one, labeled
AU, uses the Argonne AV4$'$ of Ref.~\cite{wir2002} for the $NN$ interaction,
and the Usmani potential of Ref.~\cite{usm2008} for the
$\Lambda N$ interaction (see also Ref.~\cite{fer2017}).

\subsection{Convergence study}
\label{subsec:conv}
We recall that the wave function is written as in Eq.~(\ref{eq:wf}),
and that, since we are using central potentials, only the first channel
of Table~\ref{tab:channel} is considered, as for instance in
Ref.~\cite{kie2008}. Therefore we need to study
the convergence of our results on $G^{max}$ and $N_{max}$.
Furthermore, we introduce the value of $j$ as
${\vec{j}}={\vec{\ell}}_2+{\vec{S}}$,
$\ell_2$ and $S$ being the orbital angular momentum and the spin
of the pair $ij$ on which the potential acts. This allows to set
up the theoretical framework also in the case of
projecting potentials. Therefore, we will study the convergence of our
results also on the maximum value of $j$, called $j_{max}$.
Finally, the radial function written as in Eq.~(\ref{eq:laguerre}),
presents a non-linear parameter $\gamma$, for which we need to find a range
of values such that the binding energy is stable. Note that in these
convergence studies we have used $m_n=m_p$.
%

\begin{figure}[h!]
  \begin{center}
    \includegraphics[width=1.\textwidth,height=0.7\textheight]{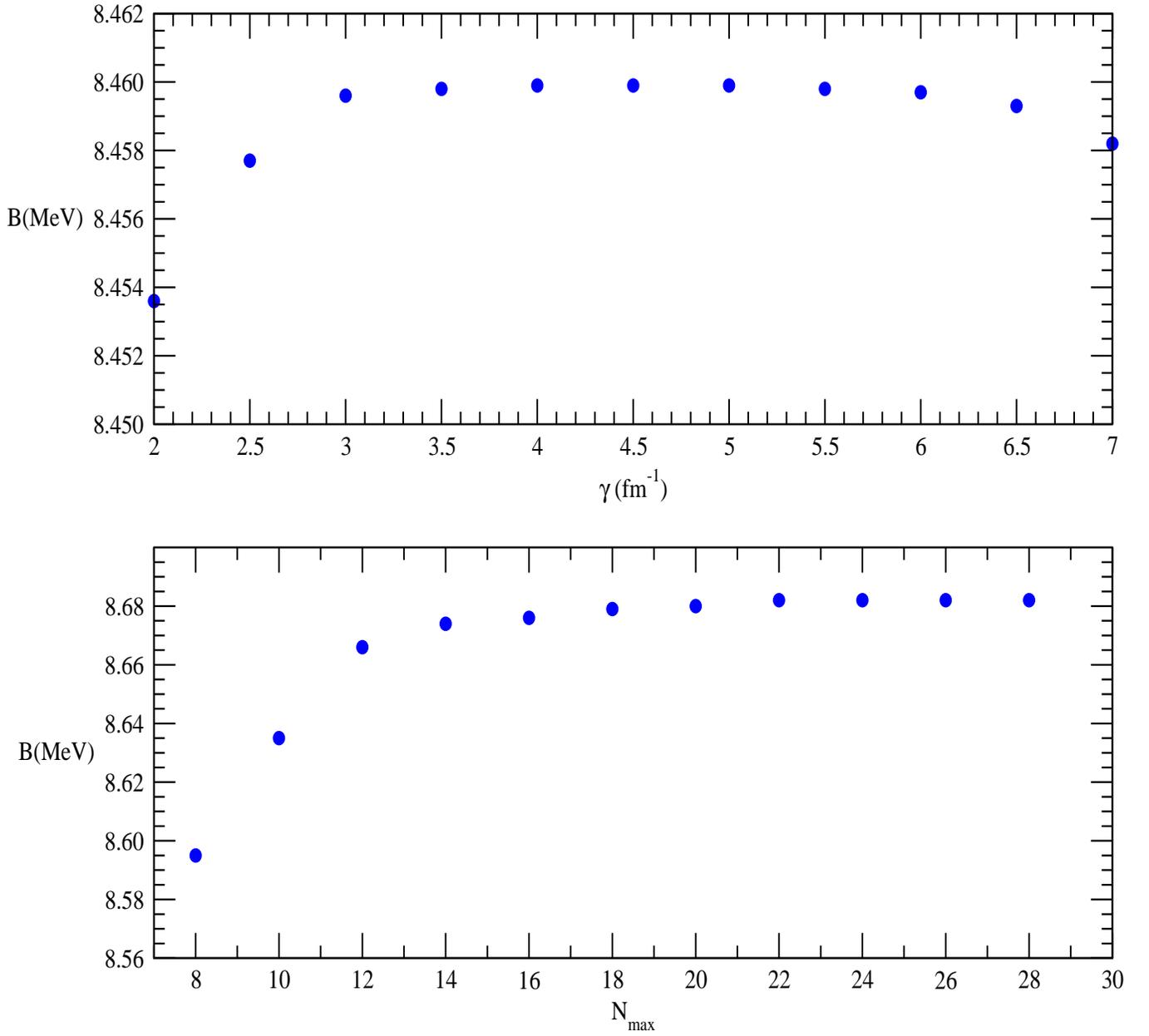}
\end{center}
\caption{Top panel: The binding energy $B$ (in MeV) as function of the
  parameter $\gamma$ (in fm$^{-1}$) for the Volkov potential
  model~\cite{vol1965}, with $G^{max}=20$, $j_{max}=6$ and
  $N_{max}=16$, and using $m_n=m_p$. Bottom panel:
  The binding energy $B$ (in MeV) as function of the parameter
  $N_{max}$ for the AV4$'$ potential model~\cite{wir2002},
  with $G^{max}=20$, $j_{max}=8$ and $\gamma=4$ fm$^{-1}$,
  and using $m_n=m_p$.} \label{fig:gamma_nnl}
\end{figure}

We start by considering the parameter $\gamma$. 
The behaviour of the binding energy as a function of $\gamma$ is shown for the
Volkov potential in the top panel of
Fig.~\ref{fig:gamma_nnl}. We mention here that for all the
other potential models we have considered, the results are similar.
The other parameters were kept constant, i.e.\
$G^{max}=20$, $N_{max}=16$ and $j_{max}=6$. The particular
dependence on $\gamma$ of the binding energy, that increases for low values
of ${\gamma}$, is constant for some central values, and decreases again for
large values of $\gamma$, allows to determine a so-called plateau, and the
optimal value for $\gamma$ has to be chosen on this plateau.
Alternatively, we can chose $\gamma$ such that for a given $N_{max}$
the binding energy is maximum. A choice of $\gamma$ outside the
plateau would require just a larger value of $N_{max}$.
To be noticed that this particular choice of $\gamma$ is not universal.
As an example, in the ``standard'' HH method, $\gamma=2.5-4.5$ fm$^{-1}$
for the AV18 potential, but much larger ($\simeq 7$ fm$^{-1}$) for the
chiral non-local potentials~\cite{kie2008}. In our case, different values
of $\gamma$ for different potentials might improve the convergence on
$N_{max}$, but not that on $j_{max}$ and $G^{max}$, determined by
the structure of the HH functions. Since, as shown below,
the convergence on $N_{max}$ is not difficult to be achieved,
we have chosen to keep $\gamma$ at a fixed value,
i.e.\ $\gamma=4$ fm$^{-1}$ for all the potentials.

In the bottom panel of Fig.~\ref{fig:gamma_nnl}
we fix $j_{max}=8$, $\gamma=4$ fm$^{-1}$ and $G^{max}=20$,
and we show the pattern of convergence for the binding energy $B$ with
respect to $N_{max}$, in the case of the Argonne AV4$'$ model. Here convergence
is reached for $N_{max}=24$, i.e.\ we have verified that,
for higher $N_{max}$ value, $B$ changes by less than 1 keV.
To be noticed that for the other potentials, convergence is already reached
for $N_{max}=16-20$.

The variation of the binding energy as a function of $j_{max}$ and $G^{max}$
depends significantly on the adopted potential model. Therefore, we need
to analyze every single case. 
As we can see from the data of Tables~\ref{tab:vol16} and~\ref{tab:min16},
the convergence on $j_{max}$ and $G^{max}$ for the Volkov and the Minnesota
potentials is really quick,
and we can reach an accuracy better than 2 keV for $j_{max}=10$ and
$G^{max}=40$.
\begin{table}[] \centering
	\begin{tabular}{cccccc}
		\hline
		\multicolumn{2}{c}{$j_{max}=6$}&\multicolumn{2}{c}{$j_{max}=8$}&\multicolumn{2}{c}{$j_{max}=10$} \\ \hline 
		$G^{max}$&$B$&$G^{max}$&$B$&$G^{max}$&$B$\\ \hline
		20&8.460&20&8.462&20&8.462\\
		30&8.461&30&8.464&30&8.464\\
		40&8.461&40&8.464&40&8.465\\
		\hline 
	\end{tabular}
	\caption{The $^3$H binding energy $B$ (in MeV) calculated
          with the Volkov potential model~\cite{vol1965},
          using $m_n=m_p$, $N_{max}=16$ and $\gamma=4$ fm$^{-1}$,
        as function of $j_{max}$ and $G^{max}$.}
	\label{tab:vol16}
\end{table}
\begin{table}[] \centering
	\begin{tabular}{cccccc} 
		\hline
		\multicolumn{2}{c}{$j_{max}=6$}&\multicolumn{2}{c}{$j_{max}=8$}&\multicolumn{2}{c}{$j_{max}=10$} \\ \hline 
		$G^{max}$&$B$&$G^{max}$&$B$&$G^{max}$&$B$\\ \hline
		20&8.376&20&8.381&20&8.381\\
		30&8.378&30&8.383&30&8.385\\
		40&8.378&40&8.383&40&8.385\\
		\hline
	\end{tabular}
	\caption{Same as Table~\ref{tab:vol16} but for
          the Minnesota potential model~\cite{tho1977}.}
	\label{tab:min16}
\end{table}
On the other hand, in the case of the Afnan-Tang potential,
we need to go up to $j_{max}=14$ and $G^{max}=50$,
in order to get a total accuracy of our results of about 2 keV
(1 keV is due to the dependence on $N_{max}$).
This can be seen by inspection of
Table~\ref{tab:afn16}. The Malfliet-Tjon potential model implies
a convergence even slower of the expansion, and we have to go up to
$G^{max}=90$  and $j_{max}=22$, to get an uncertainty of about 3 keV,
as shown in Table~\ref{tab:mal16}. In fact, being a sum of Yukawa functions,
the Malfliet-Tjon potential model is quite difficult to be treated also with
the ``standard'' symmetrized HH method~\cite{kie2008}. 
%
\begin{table}[] \centering
	\begin{tabular}{cccccccccc}
		\hline
		\multicolumn{2}{c}{$j_{max}=6$}&
                \multicolumn{2}{c}{$j_{max}=8$}&
                \multicolumn{2}{c}{$j_{max}=10$}&
                \multicolumn{2}{c}{$j_{max}=12$}&
                \multicolumn{2}{c}{$j_{max}=14$} \\ \hline 
		$G^{max}$&$B$&$G^{max}$&$B$&$G^{max}$&$B$&$G^{max}$&$B$&
                $G^{max}$&$B$\\ \hline
		20&6.567&20&6.617&20& 6.619&20&6.619&20&6.619\\
		30&6.605&30&6.664&30&6.682&30&6.688&30&6.688\\
		40&6.608&40&6.668&40&6.687&40&6.693&40&6.695\\
		50&6.608&50&6.668&50&6.687&50&6.694&50&6.696\\
		\hline 
	\end{tabular}
	\caption{Same as Table~\ref{tab:vol16} but for the
          Afnan-Tang potential model~\cite{afn1968}.}
	\label{tab:afn16}
\end{table}
\begin{table}[] \centering
	\begin{tabular}{cccccccccc}
		\hline
		\multicolumn{2}{c}{$j_{max}=10$}&
                \multicolumn{2}{c}{$j_{max}=14$}&
                \multicolumn{2}{c}{$j_{max}=18$}&
                \multicolumn{2}{c}{$j_{max}=20$}&
                \multicolumn{2}{c}{$j_{max}=22$}\\ \hline 
		$G^{max}$&$B$&$G^{max}$&$B$&$G^{max}$&$B$&$G^{max}$&$B$&
                $G^{max}$&$B$\\ \hline
		20&7.943&20&7.943&20&7.943&20&7.943&20&7.943\\
		30&8.155&30&8.179&30&8.179&30&8.179&30&8.179\\
		40&8.182&40&8.222&40&8.229&40&8.229&40&8.229\\
		50&8.190&50&8.231&50&8.241&50&8.243&50&8.243\\
		60&8.192&60&8.234&60&8.244&60&8.246&60&8.247\\
		70&8.193&70&8.235&70&8.245&70&8.248&70&8.249\\
		80&8.194&80&8.235&80&8.246&80&8.248&80&8.249\\
		90&8.194&90&8.235&90&8.246&90&8.248&90&8.250\\
		\hline 
	\end{tabular}
	\caption{Same as Table~\ref{tab:vol16} but for the
          Malfliet-Tjon potential model~\cite{mal1969}.}
	\label{tab:mal16}
\end{table}
%

In Tables~\ref{tab:av4g} and~\ref{tab:av4j}, we show the convergence study
for the AV4$'$, which is the most realistic  potential model used here
for the $A=3$ nuclear systems. As we can see by
inspection of the tables, in order to reach an accuracy of about 3 keV,
we have to push the calculation up to $G^{max}= 80$, $N_{max}= 24$ and
$j_{max}= 20$. Our final result of $B=8.991$ MeV, though, agrees
well with the one of Ref.~\cite{HHsym}, obtained with the ``standard''
symmetrized HH method, for which $B=8.992$ MeV.
%
\begin{table}[] \centering
	\begin{tabular}{lcccccccc} 
		\hline
		&\multicolumn{2}{c}{$j_{max}=10$}&\multicolumn{2}{c}{$j_{max}=12$}&\multicolumn{2}{c}{$j_{max}=14$}&\multicolumn{2}{c}{$j_{max}=16$} \\ \hline 
		$N_{max}=16$&$G^{max}$&$B$&$G^{max}$&$B$&$G^{max}$&$B$&$G^{max}$&$B$\\ \hline
		&20&8.682&20&8.682&20&8.682&20&8.682\\
		&40&8.923&40&8.956&40&8.970&40&8.975\\
		&60&8.927&60&8.960&60&8.975&60&8.981\\
		&80&8.927&80&8.960&80&8.975&80&8.981\\
		\hline 
		$N_{max}=20$&$G^{max}$&$B$&$G^{max}$&$B$&$G^{max}$&$B$&$G^{max}$&$B$\\ \hline
		&20&8.686&20&8.686&20&8.686&20&8.686\\
		&40&8.927&40&8.961&40&8.974&40&8.979\\
		&60&8.931&60&8.964&60&8.977&60&8.985\\
		&80&8.931&80&8.964&80&8.977&80&8.985\\
		\hline
		$N_{max}=24$&$G^{max}$&$B$&$G^{max}$&$B$&$G^{max}$&$B$&$G^{max}$&$B$\\ \hline
		&20&8.687&20&8.687&20&8.687&20&8.687\\
		&40&8.928&40&8.962&40&8.975&40&8.980\\
		&60&8.932&60&8.965&60&8.980&60&8.986\\
		&80&8.933&80&8.966&80&8.981&80&8.987\\
		\hline 
	\end{tabular}
	\caption{The $^3$H binding energy $B$ (in MeV) calculated with the
          AV4$'$ potential model~\cite{wir2002}
          as function of $N_{max}$, $j_{max}$ and $G^{max}$, using $m_n=m_p$ and
          $\gamma=4$ fm$^{-1}$.}
	\label{tab:av4g}
\end{table}
\begin{table}[] \centering
	\begin{tabular}{ccccccc}
		\hline 
		$j_{max}$&10&12&14&16&18&20\\ \hline
		$B$&8.932&8.965&8.980&8.986&8.989&8.991\\
		\hline 
	\end{tabular}
	\caption{The $^3$H binding energy $B$ (in MeV) calculated with the
          AV4$'$ potential model~\cite{wir2002}, using $m_n=m_p$, $G^{max}=60$,
          $N_{max}=24$ and $\gamma=4$ fm$^{-1}$.}
	\label{tab:av4j}
\end{table}

The results for the binding energy of $^3$H and $^3$He with the different
potentials will be summarized in the next Subsection.
\subsection{The $^3$H and $^3$He systems}
\label{subsec:3h-3he}

Having verified that our method can be pushed up to convergence,
we present in  the third column of Table~\ref{tab:ris1} the results
for the $^3$H binding energy with all the different potential models,
obtained still keeping $m_p=m_n$. The results are compared with those
present in the literature, finding an overall nice agreement.
\begin{table}[] \centering
	\begin{tabular}{llccccc} 
		\hline 
		Potential model&literature&$B(m_p=m_n)$&
                $B_{^3{\rm H}}$&$B_{^3{\rm He}}$&$\Delta B$&$BC_{^3{\rm He}}$ \\
		\hline
		Volkov&8.465~\cite{kie2008}&8.465&8.470&8.459&0.011&7.754\\
		Afnan-Tang&6.698~\cite{HHsym}&6.697&6.704&6.690&0.014&5.990\\
		Malfliet-Tjon&8.253~\cite{kie2008}&8.250&8.257&8.243&0.014&7.516\\
		Minnesota&8.386~\cite{kie2008}&8.385&8.389&8.381&0.008&7.706\\
		AV4$'$&8.992~\cite{HHsym}&8.991&8.998&8.984&0.014&8.272\\
		\hline
	\end{tabular}
	\caption{The $^3$H binding energy obtained using $m_n=m_p$
          ($B(m_p=m_n)$), the $^3$H and $^3$He binding energies
          calculated taking into account the difference of masses
          but no Coulomb interaction in $^3$He
          ($B_{^3{\rm H}}$ and $B_{^3{\rm He}}$), the difference
          $\Delta B=B_{^3{\rm H}}-B_{^3{\rm He}}$, and
          the $^3$He binding energy calculated including also the
          (point) Coulomb
          interaction ($BC_{^3{\rm He}}$). All the values are given in MeV.
          The results present in the literature for $B(m_p=m_n)$
          are also listed with the corresponding references.}
	\label{tab:ris1}
\end{table}

We now turn our attention to the $^3$H and $^3$He nuclei,
considering them as made of different mass particles. Therefore,
we impose $m_p\ne m_n$ and we calculate
the $^3$H and $^3$He binding energy and the difference of these
binding energies, i.e.\
\begin{equation} 
  \Delta B=B_{^3\textrm{H}}-B_{^3\textrm{He}} \ .
  \label{eq:deltab5}
\end{equation}
To be noticed that we have not yet included the effect of the (point)
Coulomb interaction. The results are
listed in Table~\ref{tab:ris1}. By inspection of the table,
we can see that $\Delta B$ is not the same for all the potential models.
In fact, while for the spin-independent  Afnan-Tang and Malfliet-Tjon
central potentials, and for the spin-dependent AV4$'$ potential,
$\Delta B=14$ keV, for the Volkov and the Minnesota potential we find a
smaller value. In all cases, though, we have verified that
$\Delta B$ is equally distributed, i.e.\ we have verified that
\begin{equation}
  B_{m_n=m_p}=B_{^3\textrm{H}}-\frac{\Delta B}{2} = B_{^3\textrm{He}} +
  \frac{\Delta B}{2} \ ,
  \label{eq:deltab}
\end{equation}
as can be seen from Table~\ref{tab:ris1}.
We would like to remark that in the NSHH method, the inclusion of the
difference of masses is quite straightforward, and $\Delta B$
can be calculated ``exactly''. This is not so trivial within the
symmetrized HH method. Furthermore, we compare our results with those
of Ref.~\cite{per}, where $\Delta B$
was calculated within the Faddeev equation method using 
realistic Argonne AV18~\cite{wir1995} potential, and it was found
$\Delta B=14$ keV, in perfect agreement with our AV4$'$ result.

In order to test our results for $\Delta B$, we try to
get a perturbative rough estimate of
$\Delta B$, proceeding as follows: since
the neutron-proton difference of mass
$\Delta m= m_n-m_p=1.2934$ MeV 
is about three orders of magnitude smaller than their
average mass $m=({m_n+m_p})/{2}=938.9187$ MeV,
we can assume also $\Delta B$ to be small. Furthermore, we suppose
the potential to be insensitive to $\Delta m$, and we consider only
the kinetic energy. 
In the center of mass frame, the kinetic energy operator
can be cast in the form
\begin{equation} 
  T= \sum_{i=1}^3 \frac{\textbf{p}_i^2}{2m_i}=
  \frac{\textbf{p}_1^2+\textbf{p}_2^2}{2m_e}+ \frac{\textbf{p}_3^2}{2m_d}\ ,
  \label{eq:t5}
\end{equation}
where $m_e$ stands for the mass of the two equal particles, i.e.\
$m_n$ for $^3$H and $m_p$ for $^3$He, and $m_d$ is the mass of the third
particle, different from the previous ones. 
By defining $E=\left<H\right>=\left<T+V\right>$, where $\left<H\right>$
is the average value of the Hamiltonian $H$, we obtain
\begin{eqnarray}
  \frac{\partial E}{\partial m_e} &=&
  \left< \frac{\partial H}{\partial m_e}\right>=
  \left< \frac{\partial T}{\partial m_e}\right> =
- \frac{\left<2 T_e\right>}{m_e} \ , \label{eq:E1}\\ 
  \frac{\partial E}{\partial m_d} &=&
  \left< \frac{\partial H}{\partial m_d}\right>=
  \left< \frac{\partial T}{\partial m_d}\right> =
- \frac{\left< T_d\right>}{m_d} \ , \label{eq:E2}
\end{eqnarray}
where we have indicated $\left< T_{e/d}\right>\approx{\bf p}_i^2/(2m_{e/d})$.
Moreover, we define the
proton and neutron mass difference $\Delta m_{p/n}$ as
\begin{eqnarray}
&&\Delta m_p \equiv m_p-m=-\frac{\Delta m}{2} \ , \label{eq:M1} \\
&&\Delta m_n \equiv m_n-m= \ \ \frac{\Delta m}{2}  \ , \label{eq:M2}
\end{eqnarray}
and the $^3$He and $^3$H binding energy difference
$\Delta B_{^3\textrm{He}/^3\textrm{H}}$ as
\begin{eqnarray}
 \Delta B_{^3\textrm{He}}&\equiv& B_{m_n=m_p}-B_{^3\textrm{He}} \ ,
  \label{eq:deltaB3He} \\
  \Delta B_{^3\textrm{H}}&=&B_{m_n=m_p}-B_{^3\textrm{H}}  \ .
  \label{eq:deltaB3H}
\end{eqnarray}
Then using Eqs.~(\ref{eq:E1})--(\ref{eq:M2}), we obtain
\begin{eqnarray}
  \Delta B_{^3\textrm{He}}&\approx& \frac{\partial E}{\partial m_e} \Delta m_p
  + \frac{\partial E}{\partial m_d} \Delta m_n =
  \frac{\left<2T_e\right>}{m_e} \frac{\Delta m}{2} -
  \frac{\left<T_d\right>}{m_d} \frac{\Delta m}{2} \nonumber \\
&\approx& \left< 2T_e-T_d \right> \frac{\Delta m}{2m} \ .
  \label{eq:deltaB3He-1} \\
  \Delta B_{^3\textrm{H}}&\approx& \frac{\partial E}{\partial m_e} \Delta m_n
  + \frac{\partial E}{\partial m_d} \Delta m_p =
  -\frac{\left<2T_e\right>}{m_e} \frac{\Delta m}{2} +
  \frac{\left<T_d\right>}{m_d} \frac{\Delta m}{2} \nonumber \\
&\approx& -\left< 2T_e-T_d \right> \frac{\Delta m}{2m} \ .
  \label{eq:deltaB3H-1}
\end{eqnarray}
In conclusion 
\begin{equation} 
  \Delta B_{PT}\equiv B_{^3\textrm{H}}- B_{^3\textrm{He}}\approx \left<2T_e-T_d\right> \frac{\Delta m}{m} \approx \left<T\right> \frac{\Delta m}{3m} \ ,
  \label{eq:deltapt}
\end{equation}
where the last equality holds assuming that
$\left<T_e\right>=\left<T_d\right>=\left<T\right>/3$, since
the $^3$He and $^3$H have a large $S$-wave component (about 90 \%).
The results of $\left<T\right>$ and $\Delta B_{PT}$ are listed
in Table~\ref{tab:riskin}, and are compared with the
values for $\Delta B$ calculated within the NSHH and already listed
in Table~\ref{tab:ris1}. By inspection of the table
we can see an overall nice agreement between this rough estimate
and the exact calculation for all the potential models.
Only in the case of the Minnesota and AV4$'$ potentials, $\Delta B_{PT}$ is
4 and 3 keV larger than $\Delta B$, respectively.
This can be understood by noticing that these potentials are spin-dependent,
giving rise to mixed-symmetry components in the wave functions. These
components are responsible for a reduction in $\Delta B_{PT}$~\cite{fri1990},
related to the fact that the nuclear force for the $^3$S$_1$ $np$ pair is
stronger than for the $^1$S$_0$ $nn$ (or $pp$) pair. Therefore, the kinetic
energy for equal particles $\left<T_e\right>$ is less than the kinetic
energy for different particles $\left< T_d\right>$.
\begin{table}[] \centering
	\begin{tabular}{l|ccc} 
		\hline 
		Potential model&$\left<T\right>$ (MeV) &$\Delta B_{PT}$ (MeV) & $\Delta B_{NSHH}$ (MeV)\\
		\hline
		Volkov&23.798&0.011&0.011 \\
		Afnan-Tang&30.410&0.014&0.014\\
		Malfliet-Tjon&30.973&0.014&0.014\\
		Minnesota&27.216&0.012&0.008\\
		AV4$'$&37.599&0.017&0.014\\
		\hline 
	\end{tabular}
	\caption{Mean value for the kinetic energy operator
          $\left<T\right>$, $\Delta B$ estimated with the perturbative theory
          ($PT$), and $\Delta B$ calculated with the NSHH for the different
          potential models considered in this work. See text for more details.}
	\label{tab:riskin}
\end{table}
\subsection{The $^3_{\Lambda}$H hypernucleus}
\label{subsec:hyper3h}
The hypernucleus $^3_{\Lambda}$H is a bound system
composed by a neutron, a proton and the $\Lambda$ hyperon.
In order to study this system, we have considered the proton and the neutron
as reference-pair, with equal mass $m_n=m_p=m$, while the $\Lambda$
particle has been taken as the third particle with different mass.
The $\Lambda$ hyperon mass has been chosen depending on the
considered potential. We remind that we have used three different
potential models: a central spin-independent
Gaussian model~\cite{cla1985}, and two spin-dependent
central potentials, labelled MN9~\cite{phd2017} and
AU~\cite{fer2017} potentials.
Therefore, when the  $^3_{\Lambda}$H hypernucleus
has been studied using the Gaussian potential of Ref.~\cite{cla1985},
we have set $M_{\Lambda}=6/5 \ m_{N}$, accordingly. In the other two cases,
we have used $M_{\Lambda}=1115.683$ MeV. 
We first study the convergence pattern of our method, which
in the case of the Gaussian potential of Ref.~\cite{cla1985} is really
fast, with a reached accuracy of 1 keV on the binding energy already with
$N_{max}=20$, $j_{max}=10$ and $G^{max}=50$. This can be seen
directly by inspection of Tables~\ref{tab:cen} and~\ref{tab:cennnl}. 
\begin{table}[] \centering
	\begin{tabular}{cccccc} 
		\hline
		\multicolumn{2}{c}{$j_{max}=6$}&
                \multicolumn{2}{c}{$j_{max}=8$}&
                \multicolumn{2}{c}{$j_{max}=10$} \\ \hline
		$G^{max}$&$B$&$G^{max}$&$B$&$G^{max}$&$B$\\ \hline
		0&0.510&0&0.510&0&0.510\\
		2&1.070&2&1.070&2&1.070\\
		4&1.776&4&1.776&4&1.776\\
		6&2.211&6&2.211&6&2.211\\
		8&2.371&8&2.371&8&2.371\\
		10&2.476&10&2.476&10&2.476\\
		12&2.551&12&2.551&12&2.551\\
		20&2.659&20&2.660&20&2.660\\
		30&2.692&30&2.693&30&2.693\\
		40&2.700&40&2.701&40&2.701\\
		50&2.702&50&2.703&50&2.703\\
		\hline 
	\end{tabular}
	\caption{The $^3_{\Lambda}$H binding energy $B$ (in MeV)
          as function of $j_{max}$ and $G^{max}$, calculated with
          the Gaussian potential model of Ref.~\cite{cla1985}, using
          $N_{max}=20$ and $\gamma=4$ fm$^{-1}$.}
	\label{tab:cen}
\end{table}
\begin{table}[] \centering
	\begin{tabular}{ccccc}
		\hline 
		$N_{max}$&8&12&16&20\\ \hline
		$B$&2.552&2.651&2.660&2.660\\
		\hline 
	\end{tabular}
	\caption{The $^3_{\Lambda}$H binding energy $B$ (in MeV)
          as function of $N_{max}$, calculated with the the Gaussian
          potential model of Ref.~\cite{cla1985}, using $G^{max}=20$,
          $j_{max}=8$ and $\gamma=4$ fm$^{-1}$.}
	\label{tab:cennnl}
\end{table}

The convergence pattern in the case of the spin-dependent central
MN9 and AU potentials has been found quite slower. This is shown
in Tables~\ref{tab:minh} and ~\ref{tab:usm}, respectively.
\begin{table}[] \centering
	\begin{tabular}{lcccccccc} 
		\hline
		&\multicolumn{2}{c}{$j_{max}=6$}&
                \multicolumn{2}{c}{$j_{max}=10$}&
                \multicolumn{2}{c}{$j_{max}=12$}&
                \multicolumn{2}{c}{$j_{max}=14$} \\ \hline 
		$N_{max}=16$&$G^{max}$&$B$&$G^{max}$&$B$&$G^{max}$&$B$&
                $G^{max}$&$B$ \\ \hline
		&50&2.174&50&2.201&50&2.205&50&2.207\\
		&60&2.178&60&2.205&60&2.209&60&2.211\\
		&70&2.181&70&2.207&70&2.211&70&2.213\\
		&80&2.181&80&2.208&80&2.212&80&2.214\\
		&90&2.182&90&2.208&90&2.212&90&2.215\\
		&100&2.182&100&2.208&100&2.212&100&2.215\\
		\hline 
		$N_{max}=20$&$G^{max}$&$B$&$G^{max}$&$B$&$G^{max}$&$B$&
                $G^{max}$&$B$\\ \hline
		&50&2.206&50&2.232&50&2.236&50&2.239\\
		&60&2.213&60&2.238&60&2.242&60&2.245\\
		&70&2.216&70&2.241&70&2.246&70&2.248\\
		&80&2.219&80&2.243&80&2.248&80&2.250\\
		&90&2.220&90&2.244&90&2.249&90&2.251\\
		&100&2.220&100&2.244&100&2.249&100&2.252\\
		\hline
		$N_{max}=24$&$G^{max}$&$B$&$G^{max}$&$B$&$G^{max}$&$B$&
                $G^{max}$&$B$\\ \hline
		&50&2.219&50&2.243&50&2.247&50&2.250\\
		&60&2.227&60&2.251&60&2.255&60&2.257\\
		&70&2.232&70&2.255&70&2.259&70&2.261\\
		&80&2.235&80&2.257&80&2.261&80&2.264\\
		&90&2.236&90&2.259&90&2.263&90&2.266\\
		&100&2.236&100&2.260&100&2.264&100&2.267\\
		\hline
		$N_{max}=28$&$G^{max}$&$B$&$G^{max}$&$B$&$G^{max}$&
                $B$&$G^{max}$&$B$\\ \hline
		&50&2.222&50&2.248&50&2.252&50&2.255\\
		&60&2.233&60&2.257&60&2.261&60&2.264\\
		&70&2.240&70&2.263&70&2.267&70&2.269\\
		&80&2.244&80&2.266&80&2.270&80&2.272\\
		&90&2.246&90&2.268&90&2.272&90&2.274\\
		&100&2.248&100&2.269&100&2.273&100&2.276\\
		\hline
		$N_{max}=32$&$G^{max}$&$B$&$G^{max}$&$B$&$G^{max}$&
                $B$&$G^{max}$&$B$\\ \hline
		&50&2.225&50&2.249&50&2.253&50&2.256\\
		&60&2.236&60&2.258&60&2.262&60&2.265\\
		&70&2.243&70&2.264&70&2.268&70&2.272\\
		&80&2.247&80&2.268&80&2.273&80&2.275\\
		&90&2.250&90&2.272&90&2.275&90&2.277\\
		&100&2.252&100&2.273&100&2.276&100&2.279\\
		\hline
		$N_{max}=34$&$G^{max}$&$B$&$G^{max}$&$B$&$G^{max}$&
                $B$&$G^{max}$&$B$\\ \hline
		&50&2.225&50&2.249&50&2.253&50&2.256\\
		&60&2.237&60&2.259&60&2.263&60&2.266\\
		&70&2.244&70&2.265&70&2.269&70&2.272\\
		&80&2.248&80&2.269&80&2.274&80&2.276\\
		&90&2.251&90&2.273&90&2.276&90&2.278\\
		&100&2.253&100&2.274&100&2.277&100&2.280\\
		\hline
	\end{tabular}
	\caption{The $^3_{\Lambda}$H binding energy $B$ (in MeV)
          as function of $G^{max}$, $j_{max}$ and $N_{max}$, calculated
          with the MN9 potential model of Ref.~\cite{phd2017},
          using $\gamma=4$ fm$^{-1}$.}
	\label{tab:minh}
\end{table}
\begin{table}[] \centering
	\begin{tabular}{llccccccccc}
		\hline
&		\multicolumn{2}{c}{$j_{max}=6$}&\multicolumn{2}{c}{$j_{max}=10$}&\multicolumn{2}{c}{$j_{max}=12$}&\multicolumn{2}{c}{$j_{max}=14$}&\multicolumn{2}{c}{$j_{max}=16$} \\ \hline 
$N_{max}=16$ & $G^{max}$&$B$&$G^{max}$&$B$&$G^{max}$&$B$&$G^{max}$&$B$&$G^{max}$&$B$ \\ \hline
&		20&1.436&20&1.508&20&1.521&20&1.521&20&1.521\\
&		30&1.924&30&2.037&30&2.051&30&2.056&30&2.056\\
&		40&2.137&40&2.243&40&2.258&40&2.267&40&2.270\\
&		50&2.250&50&2.349&50&2.363&50&2.372&50&2.375\\
&		60&2.314&60&2.408&60&2.421&60&2.430&60&2.433\\
&		70&2.355&70&2.445&70&2.455&70&2.461&70&2.467\\
&		80&2.379&80&2.466&80&2.476&80&2.484&80&2.485\\
&		90&2.394&90&2.479&90&2.489&90&2.497&90&2.499\\
&		100&2.402&100&2.486&100&2.498&100&2.503&100&2.506\\
&		110&2.406&110&2.489&110&2.504&110&2.510&110&2.514\\
&		120&2.408&120&2.491&120&2.507&120&2.513&120&2.518\\
&		130&2.409&130&2.492&130&2.509&130&2.515&130&2.520\\
&		140&2.409&140&2.492&140&2.510&140&2.516&140&2.521\\
		\hline 
$N_{max}=20$ & $G^{max}$&$B$&$G^{max}$&$B$&$G^{max}$&$B$&$G^{max}$&$B$&$G^{max}$&$B$ \\ \hline
&		20&1.438&20&1.522&20&1.522&20&1.522&20&1.522\\
&		30&1.925&30&2.038&30&2.053&30&2.057&30&2.057\\
&		40&2.139&40&2.245&40&2.259&40&2.268&40&2.271\\
&		50&2.252&50&2.351&50&2.364&50&2.373&50&2.376\\
&		60&2.317&60&2.409&60&2.423&60&2.431&60&2.435\\
&		70&2.357&70&2.446&70&2.458&70&2.466&70&2.469\\
&		80&2.384&80&2.470&80&2.480&80&2.488&80&2.491\\
&		90&2.402&90&2.485&90&2.495&90&2.501&90&2.505\\
&		100&2.413&100&2.494&100&2.504&100&2.512&100&2.515\\
&		110&2.421&110&2.501&110&2.510&110&2.518&110&2.521\\
&		120&2.426&120&2.505&120&2.514&120&2.521&120&2.525\\
&		130&2.429&130&2.507&130&2.516&130&2.524&130&2.527\\
&		140&2.430&140&2.508&140&2.517&140&2.526&140&2.528\\
		\hline
$N_{max}=24$ & $G^{max}$&$B$&$G^{max}$&$B$&$G^{max}$&$B$&$G^{max}$&$B$&$G^{max}$&$B$ \\ \hline
&		20&1.438&20&1.523&20&1.523&20&1.524&20&1.524\\
&		30&1.926&30&2.039&30&2.053&30&2.057&30&2.057\\
&		40&2.139&40&2.245&40&2.259&40&2.268&40&2.271\\
&		50&2.252&50&2.351&50&2.364&50&2.373&50&2.377\\
&		60&2.317&60&2.410&60&2.423&60&2.432&60&2.435\\
&		70&2.357&70&2.446&70&2.458&70&2.467&70&2.470\\
&		80&2.385&80&2.471&80&2.481&80&2.488&80&2.492\\
&		90&2.402&90&2.485&90&2.496&90&2.502&90&2.506\\
&		100&2.414&100&2.494&100&2.505&100&2.513&100&2.516\\
&		110&2.423&110&2.502&110&2.511&110&2.519&110&2.522\\
&		120&2.429&120&2.507&120&2.516&120&2.523&120&2.526\\
&		130&2.433&130&2.510&130&2.519&130&2.527&130&2.530\\
&		140&2.437&140&2.511&140&2.521&140&2.530&140&2.532\\
		\hline
	\end{tabular}
	\caption{Same as Table~\ref{tab:minh} but using the
          AU potential model of Ref.~\cite{fer2017} for $N_{max}=16,20,24$.}
	\label{tab:usm}
\end{table}
By inspection of Table~\ref{tab:minh}, we can conclude that
$B=2.280$ MeV, with an accuracy of about 3 keV, obtained
with $G^{max}=100$, $N_{max}=34$, and $j_{max}=14$.
By inspection of Table~\ref{tab:usm}, $B=2.532$ MeV, with an accuracy
of about 4 keV, going up to $G^{max}=140$, $N_{max}=24$
and $j_{max}=16$.

The results obtained with our method for the three potential models
considered in this work are compared with those present in the
literature~\cite{cla1985,phd2017,fer2017} in Table~\ref{ris2},
finding a very nice agreement,
within the reached accuracy.

\begin{table}[] \centering
	\begin{tabular}{lcc} 
		\hline 
		Potential model&$B$&literature \\
		\hline
		Gaussian& \ 2.703&2.71~\cite{cla1985}\\
		MN9&2.280&2.27~\cite{phd2017}\\
		AU&2.532&2.530~\cite{fer2017}\\
		\hline 
	\end{tabular}
	\caption{
          The $^3_{\Lambda}$H binding energy $B$ (in MeV)
          obtained in the present work is compared with the results
          present in the literature.}
	\label{ris2}
\end{table}

\section{Conclusions and outlook}
\label{sec:conc}
In this work we present a study of the bound state of a three-body system,
composed of different particles, by means of the NSHH method.
The method has been reviewed in Section~\ref{sec:form}.
In order to verify its validity, we have started by considering a system of
three equal-mass nucleons interacting  via different
central potential models, three spin-independent
and two spin-dependent. We have studied the convergence
pattern, and we have compared our results at convergence with those
present in the literature, finding an overall nice agreement.
Then, we have switched on the difference of mass between protons and neutrons
and we have calculated the difference of binding energy $\Delta B$
due to the difference between the neutron and proton masses. We have found that
$\Delta B$ depends on the considered potential model, but is
always symmetrically distributed (see Eq.~(\ref{eq:deltab})).

Finally we have implemented our method for the $^3_{\Lambda}$H hypernucleus,
studied with three different potentials, i.e.
the Gaussian potential of Ref.~\cite{cla1985}, for which we have found a fast convergence of the NSHH method, the MN9 and the AU potentials
of Ref.~\cite{phd2017}, for which the convergence is much slower.
In these last two cases, in particular, we had found
necessary to include a large number of the HH basis
(46104 for the MN9 and 52704 for the AU potentials),
but the agreement with the results in the literature has been found
quite nice.
To be noticed that we have included only two-body interactions,
and therefore a comparison with the experimental data is
meaningless.

In conclusion, we believe that we have proven the NSHH method
to be a good choice for studying three-body systems composed of
two equal mass particles, different from the mass of the third particle.
Besides $^3$H, $^3$He, and $_\Lambda^3$H, several other nuclear systems
can be viewed as three-body systems of different masses. This applies
in all cases where a strong clusterization is present, as in the 
case of $^6$He and $^6$Li nuclei, seen as $NN\alpha$ , or the $^9$Be
and $^9$B, seen as a $\alpha\alpha N$ three-body
systems. Furthermore, taking advantage of the versatility
of the HH method also for scattering systems, the NSHH approach
could be extended as well to scattering problems.
Work along these lines are currently underway.

\section*{Acknowledgments}
The authors would like to thank Dr. F.\ Ferrari Ruffino for useful
discussions. Computational resources provided by the INFN-Pisa Computer Center are gratefully acknowledged.

\section{Appendix: The Transformation coefficients}
\label{app}

Let us start by writing Eq.~(\ref{eq:a}) as
\begin{equation}
  a_{\ell_1,\ell_2,n,\ell_1',\ell_2',n'}^{(p \rightarrow p'),G,\Lambda}=
  \int d \Omega^{(p')} \
       [H_{[\ell_1',\ell_2',n',\Lambda \Lambda_z]}(\Omega^{(p')})]^\dagger \
       {H_{[\ell_1,\ell_2,n,\Lambda \Lambda_z]}}(\Omega^{(p)}) \ , 	
\label{eq:aa}
\end{equation} 
where $\ell_i$ ($\ell'_i$) is the orbital angular momentum
associated with the Jacobi
coordinate $\textbf{y}_i^{(p)}$ ($\textbf{y}_i^{(p')}$). 
It can be demonstrated by direct calculation and exploiting the
spherical harmonics proprieties that
\begin{eqnarray}
  a_{\ell_1,\ell_2,n,\ell_1',\ell_2',n'}^{(p \rightarrow p'),G,L} &=& N_{n'}^{\ell_1', \ell_2'}
  N_{n}^{\ell_1, \ell_2} \, \frac{1}{2} \,
  \int_{0}^{\frac{\pi}{2}} d \phi \int_{-1}^{1} d \mu \
  (\cos{\phi}^{(p')})^{2+\ell_2'} (\sin{\phi}^{(p')})^{2+\ell_1'}  \nonumber \\
&\times& P_{n'}^{\ell'_1+1/2,\ell'_2+1/2}(\cos{2\phi}^{(p')})
P_{n}^{\ell_1+1/2,\ell_2+1/2}(\cos{2\phi}^{(p)})  \nonumber\\
&\times& \sum_{\lambda,\lambda_1,\lambda_2}
C^{(p),(p')}_{\ell_1,\ell_2,\lambda_1,\lambda_2}(\sin{\phi}^{(p')},\cos{\phi}^{(p')})
P_{\lambda}(\mu)  \nonumber\\
&\times& (-)^{\Lambda+\lambda_2+\ell_2'}
(2 \lambda +1) \hat{\ell'}_1\hat{\ell'}_2\hat{\lambda}_1\hat{\lambda}_2
\nonumber 
\\
&\times& 
\begin{Bmatrix}
\ell'_1 & \ell'_2 & \Lambda \\
\lambda_2 & \lambda_1 & \lambda \\
\end{Bmatrix} 
\begin{pmatrix}
\ell'_1 & \lambda_1 & \lambda \\
0&0&0
\end{pmatrix}
\begin{pmatrix}
\ell'_2 & \lambda_2 & \lambda \\
0&0&0
\end{pmatrix} \ .
\label{eq:TC}
\end{eqnarray}
Here the curly brackets indicate the $6j$ Wigner coefficients,
and the coefficients
$C^{(p),(p')}_{\ell_1,\ell_2,\ell_1',\ell_2'} (\sin{\phi}^{(p')},\cos{\phi}^{(p')})$
are defined as
\begin{eqnarray}
  C^{(p),(p')}_{\ell_1,\ell_2,\ell_1',\ell_2'} (\sin{\phi}^{(p')},\cos{\phi}^{(p')})
  &=& \sum_{\lambda_1+\lambda_2=\ell_1}
  \sum_{\lambda'_1+\lambda'_2=\ell_2} (\sin{\phi}^{(p')})^{\lambda_1+\lambda'_1}
  (\cos{\phi}^{(p')})^{\lambda_2+\lambda'_2} \nonumber\\
  &\times&  (\alpha_{11(p')}^{(p)})^{\lambda_1}(\alpha_{12(p')}^{(p)})^{\lambda_2}(\alpha_{21(p')}^{(p)})^{\lambda'_1}
  (\alpha_{22(p')}^{(p)})^{\lambda'_2}   \nonumber \\ 
  &\times& (-)^{\lambda_1+\lambda_2+\lambda'_1+\lambda'_2} \
  D_{\ell_1,\lambda_1,\lambda_2} D_{\ell_2,\lambda'_1,\lambda'_2} 
\nonumber \\
&\times& \hat{\ell_1}\hat{\ell_2}\hat{\ell'_1}\hat{\ell'_2}
\hat{\lambda}_1\hat{\lambda}_2\hat{\lambda}'_1\hat{\lambda}'_2
\begin{pmatrix}
\lambda_1&\lambda'_1 & \ell'_1 \\
0&0&0
\end{pmatrix} \nonumber \\
&\times&
\begin{pmatrix}
\lambda_2&\lambda'_2 & \ell'_2 \\
0&0&0
\end{pmatrix} 
\begin{Bmatrix}
\lambda_1&\lambda_2 & \ell_1 \\
\lambda'_1&\lambda'_2 & \ell_2 \\
\ell'_1&\ell'_2 & \Lambda 
\end{Bmatrix} \ .
\label{eq:c}
\end{eqnarray}
In Eq.~(\ref{eq:c}) $\hat{\ell} \equiv \sqrt{2\ell+1}$, and the round
(curly) brackets denote $3j$ ($9j$) Wigner coefficients.  
The coefficients $\alpha_{ij(p')}^{(p)}$, with $ij=1,2$ are given by
		\begin{equation}\label{app11}
		\textbf{y}_i^{(p)}= \sum_{j=1}^2 \alpha_{ij(p')}^{(p)}  \textbf{y}_j^{(p')} \ ,
		\end{equation}
and depend on the (different) masses of the three particles.
and $D_{\ell,\ell_a,\ell_b}$ is defined as
\begin{equation}
  D_{\ell,\ell_a,\ell_b}=\sqrt{\frac{ (2\ell+1)!}{ (2\ell_a+1)! (2\ell_b+1)! }}
  \ .
  \label{eq:deel}
\end{equation}



\begin{thebibliography}{24}
\expandafter\ifx\csname natexlab\endcsname\relax\def\natexlab#1{#1}\fi
\expandafter\ifx\csname bibnamefont\endcsname\relax
  \def\bibnamefont#1{#1}\fi
\expandafter\ifx\csname bibfnamefont\endcsname\relax
  \def\bibfnamefont#1{#1}\fi
\expandafter\ifx\csname citenamefont\endcsname\relax
  \def\citenamefont#1{#1}\fi
\expandafter\ifx\csname url\endcsname\relax
  \def\url#1{\texttt{#1}}\fi
\expandafter\ifx\csname urlprefix\endcsname\relax\def\urlprefix{URL }\fi
\providecommand{\bibinfo}[2]{#2}
\providecommand{\eprint}[2][]{\url{#2}}

\bibitem[{\citenamefont{Kievsky et~al.}(2008)\citenamefont{Kievsky, Rosati,
  Viviani, Marcucci, and Girlanda}}]{kie2008}
\bibinfo{author}{\bibfnamefont{A.}~\bibnamefont{Kievsky}},
  \bibinfo{author}{\bibfnamefont{S.}~\bibnamefont{Rosati}},
  \bibinfo{author}{\bibfnamefont{M.}~\bibnamefont{Viviani}},
  \bibinfo{author}{\bibfnamefont{L.~E.} \bibnamefont{Marcucci}},
  \bibnamefont{and} \bibinfo{author}{\bibfnamefont{L.}~\bibnamefont{Girlanda}},
  \bibinfo{journal}{J. Phys. G} \textbf{\bibinfo{volume}{35}},
  \bibinfo{pages}{063101} (\bibinfo{year}{2008}), \eprint{0805.4688}.

\bibitem[{\citenamefont{Leidemann and Orlandini}(2013)}]{lei2013}
\bibinfo{author}{\bibfnamefont{W.}~\bibnamefont{Leidemann}} \bibnamefont{and}
  \bibinfo{author}{\bibfnamefont{G.}~\bibnamefont{Orlandini}},
  \bibinfo{journal}{Prog. Part. Nucl. Phys.} \textbf{\bibinfo{volume}{68}},
  \bibinfo{pages}{158} (\bibinfo{year}{2013}), \eprint{1204.4617}.

\bibitem[{\citenamefont{Gattobigio
  et~al.}(2009{\natexlab{a}})\citenamefont{Gattobigio, Kievsky, Viviani, and
  Barletta}}]{gat2009a}
\bibinfo{author}{\bibfnamefont{M.}~\bibnamefont{Gattobigio}},
  \bibinfo{author}{\bibfnamefont{A.}~\bibnamefont{Kievsky}},
  \bibinfo{author}{\bibfnamefont{M.}~\bibnamefont{Viviani}}, \bibnamefont{and}
  \bibinfo{author}{\bibfnamefont{P.}~\bibnamefont{Barletta}},
  \bibinfo{journal}{Phys. Rev.} \textbf{\bibinfo{volume}{A79}},
  \bibinfo{pages}{032513} (\bibinfo{year}{2009}{\natexlab{a}}),
  \eprint{0811.4259}.

\bibitem[{\citenamefont{Gattobigio
  et~al.}(2009{\natexlab{b}})\citenamefont{Gattobigio, Kievsky, Viviani, and
  Barletta}}]{gat2009b}
\bibinfo{author}{\bibfnamefont{M.}~\bibnamefont{Gattobigio}},
  \bibinfo{author}{\bibfnamefont{A.}~\bibnamefont{Kievsky}},
  \bibinfo{author}{\bibfnamefont{M.}~\bibnamefont{Viviani}}, \bibnamefont{and}
  \bibinfo{author}{\bibfnamefont{P.}~\bibnamefont{Barletta}},
  \bibinfo{journal}{Few-Body Syst.} \textbf{\bibinfo{volume}{45}},
  \bibinfo{pages}{127} (\bibinfo{year}{2009}{\natexlab{b}}), ISSN
  \bibinfo{issn}{1432-5411},
  \urlprefix\url{https://doi.org/10.1007/s00601-009-0045-4}.

\bibitem[{\citenamefont{Gattobigio et~al.}(2011)\citenamefont{Gattobigio,
  Kievsky, and Viviani}}]{gat2011}
\bibinfo{author}{\bibfnamefont{M.}~\bibnamefont{Gattobigio}},
  \bibinfo{author}{\bibfnamefont{A.}~\bibnamefont{Kievsky}}, \bibnamefont{and}
  \bibinfo{author}{\bibfnamefont{M.}~\bibnamefont{Viviani}},
  \bibinfo{journal}{Phys. Rev.} \textbf{\bibinfo{volume}{C83}},
  \bibinfo{pages}{024001} (\bibinfo{year}{2011}), \eprint{1009.3426}.

\bibitem[{\citenamefont{Deflorian et~al.}(2013)\citenamefont{Deflorian, Barnea,
  Leidemann, and Orlandini}}]{def2013}
\bibinfo{author}{\bibfnamefont{S.}~\bibnamefont{Deflorian}},
  \bibinfo{author}{\bibfnamefont{N.}~\bibnamefont{Barnea}},
  \bibinfo{author}{\bibfnamefont{W.}~\bibnamefont{Leidemann}},
  \bibnamefont{and}
  \bibinfo{author}{\bibfnamefont{G.}~\bibnamefont{Orlandini}},
  \bibinfo{journal}{Few-Body Syst.} \textbf{\bibinfo{volume}{54}},
  \bibinfo{pages}{1879} (\bibinfo{year}{2013}), \eprint{1212.5532}.

\bibitem[{\citenamefont{Deflorian et~al.}(2014)\citenamefont{Deflorian, Barnea,
  Leidemann, and Orlandini}}]{def2014}
\bibinfo{author}{\bibfnamefont{S.}~\bibnamefont{Deflorian}},
  \bibinfo{author}{\bibfnamefont{N.}~\bibnamefont{Barnea}},
  \bibinfo{author}{\bibfnamefont{W.}~\bibnamefont{Leidemann}},
  \bibnamefont{and}
  \bibinfo{author}{\bibfnamefont{G.}~\bibnamefont{Orlandini}},
  \bibinfo{journal}{Few-Body Syst.} \textbf{\bibinfo{volume}{55}},
  \bibinfo{pages}{831} (\bibinfo{year}{2014}).

\bibitem[{\citenamefont{Rosati}(2002)}]{ros}
\bibinfo{author}{\bibfnamefont{S.}~\bibnamefont{Rosati}}, in
  \emph{\bibinfo{booktitle}{Introduction to Modern Methods of Quantum Many-Body
  Theory and Their Applications, Series on Advances in Quantum Many-Body
  Theory, vol.7}}, edited by
  \bibinfo{editor}{\bibfnamefont{A.}~\bibnamefont{Fabrocini}},
  \bibinfo{editor}{\bibfnamefont{S.}~\bibnamefont{Fantoni}}, \bibnamefont{and}
  \bibinfo{editor}{\bibfnamefont{E.}~\bibnamefont{Krotscheck}}
  (\bibinfo{publisher}{World Scientific}, \bibinfo{year}{2002}), p.
  \bibinfo{pages}{339}.

\bibitem[{\citenamefont{Chen et~al.}(1986)\citenamefont{Chen, Payne, Friar, and
  Gibson}}]{che1986}
\bibinfo{author}{\bibfnamefont{C.~R.} \bibnamefont{Chen}},
  \bibinfo{author}{\bibfnamefont{G.~L.} \bibnamefont{Payne}},
  \bibinfo{author}{\bibfnamefont{J.~L.} \bibnamefont{Friar}}, \bibnamefont{and}
  \bibinfo{author}{\bibfnamefont{B.~F.} \bibnamefont{Gibson}},
  \bibinfo{journal}{Phys. Rev. C} \textbf{\bibinfo{volume}{33}},
  \bibinfo{pages}{1740} (\bibinfo{year}{1986}),
  \urlprefix\url{https://link.aps.org/doi/10.1103/PhysRevC.33.1740}.

\bibitem[{\citenamefont{Edmonds}(1974)}]{Edmonds}
\bibinfo{author}{\bibfnamefont{A.}~\bibnamefont{Edmonds}},
  \emph{\bibinfo{title}{Angular Momentum in Quantum Mechanics}}
  (\bibinfo{publisher}{Princeton University Press},
  \bibinfo{address}{Princeton, New Jersey}, \bibinfo{year}{1974}).

\bibitem[{\citenamefont{Raynal and Revai}(1970)}]{ray1970}
\bibinfo{author}{\bibfnamefont{J.}~\bibnamefont{Raynal}} \bibnamefont{and}
  \bibinfo{author}{\bibfnamefont{J.}~\bibnamefont{Revai}}, \bibinfo{journal}{Il
  Nuovo Cimento A (1965-1970)} \textbf{\bibinfo{volume}{68}},
  \bibinfo{pages}{612} (\bibinfo{year}{1970}), ISSN \bibinfo{issn}{1826-9869},
  \urlprefix\url{https://doi.org/10.1007/BF02756127}.

\bibitem[{\citenamefont{Volkov}(1965)}]{vol1965}
\bibinfo{author}{\bibfnamefont{A.}~\bibnamefont{Volkov}},
  \bibinfo{journal}{Nucl. Phys.} \textbf{\bibinfo{volume}{74}},
  \bibinfo{pages}{33} (\bibinfo{year}{1965}), ISSN \bibinfo{issn}{0029-5582},
  \urlprefix\url{http://www.sciencedirect.com/science/article/pii/0029558265902440}.

\bibitem[{\citenamefont{Afnan and Tang}(1968)}]{afn1968}
\bibinfo{author}{\bibfnamefont{I.~R.} \bibnamefont{Afnan}} \bibnamefont{and}
  \bibinfo{author}{\bibfnamefont{Y.~C.} \bibnamefont{Tang}},
  \bibinfo{journal}{Phys. Rev.} \textbf{\bibinfo{volume}{175}},
  \bibinfo{pages}{1337} (\bibinfo{year}{1968}),
  \urlprefix\url{https://link.aps.org/doi/10.1103/PhysRev.175.1337}.

\bibitem[{\citenamefont{Malfliet and Tjon}(1969)}]{mal1969}
\bibinfo{author}{\bibfnamefont{R.}~\bibnamefont{Malfliet}} \bibnamefont{and}
  \bibinfo{author}{\bibfnamefont{J.}~\bibnamefont{Tjon}},
  \bibinfo{journal}{Phys. Lett. B} \textbf{\bibinfo{volume}{30}},
  \bibinfo{pages}{293} (\bibinfo{year}{1969}), ISSN \bibinfo{issn}{0370-2693},
  \urlprefix\url{http://www.sciencedirect.com/science/article/pii/0370269369904833}.

\bibitem[{\citenamefont{Thompson et~al.}(1977)\citenamefont{Thompson, Lemere,
  and Tang}}]{tho1977}
\bibinfo{author}{\bibfnamefont{D.}~\bibnamefont{Thompson}},
  \bibinfo{author}{\bibfnamefont{M.}~\bibnamefont{Lemere}}, \bibnamefont{and}
  \bibinfo{author}{\bibfnamefont{Y.}~\bibnamefont{Tang}},
  \bibinfo{journal}{Nucl. Phys. A} \textbf{\bibinfo{volume}{286}},
  \bibinfo{pages}{53} (\bibinfo{year}{1977}), ISSN \bibinfo{issn}{0375-9474},
  \urlprefix\url{http://www.sciencedirect.com/science/article/pii/0375947477900070}.

\bibitem[{\citenamefont{Wiringa and Pieper}(2002)}]{wir2002}
\bibinfo{author}{\bibfnamefont{R.~B.} \bibnamefont{Wiringa}} \bibnamefont{and}
  \bibinfo{author}{\bibfnamefont{S.~C.} \bibnamefont{Pieper}},
  \bibinfo{journal}{Phys. Rev. Lett.} \textbf{\bibinfo{volume}{89}},
  \bibinfo{pages}{182501} (\bibinfo{year}{2002}),
  \urlprefix\url{https://link.aps.org/doi/10.1103/PhysRevLett.89.182501}.

\bibitem[{\citenamefont{Wiringa et~al.}(1995)\citenamefont{Wiringa, Stoks, and
  Schiavilla}}]{wir1995}
\bibinfo{author}{\bibfnamefont{R.~B.} \bibnamefont{Wiringa}},
  \bibinfo{author}{\bibfnamefont{V.~G.~J.} \bibnamefont{Stoks}},
  \bibnamefont{and}
  \bibinfo{author}{\bibfnamefont{R.}~\bibnamefont{Schiavilla}},
  \bibinfo{journal}{Phys. Rev. C} \textbf{\bibinfo{volume}{51}},
  \bibinfo{pages}{38} (\bibinfo{year}{1995}),
  \urlprefix\url{https://link.aps.org/doi/10.1103/PhysRevC.51.38}.

\bibitem[{\citenamefont{Clare and Levinger}(1985)}]{cla1985}
\bibinfo{author}{\bibfnamefont{R.~B.} \bibnamefont{Clare}} \bibnamefont{and}
  \bibinfo{author}{\bibfnamefont{J.~S.} \bibnamefont{Levinger}},
  \bibinfo{journal}{Phys. Rev. C} \textbf{\bibinfo{volume}{31}},
  \bibinfo{pages}{2303} (\bibinfo{year}{1985}),
  \urlprefix\url{https://link.aps.org/doi/10.1103/PhysRevC.31.2303}.

\bibitem[{\citenamefont{Ferrari~Ruffino}(2017)}]{phd2017}
\bibinfo{author}{\bibfnamefont{F.}~\bibnamefont{Ferrari~Ruffino}},
  \emph{\bibinfo{title}{Non-Symmetrized Hyperspherical Harmonics Method Applied
  to Light Hypernuclei}} (\bibinfo{publisher}{PhD Thesis, University of
  Trento}, \bibinfo{address}{Trento (Italy)}, \bibinfo{year}{2017}).

\bibitem[{\citenamefont{Usmani and Khanna}(2008)}]{usm2008}
\bibinfo{author}{\bibfnamefont{A.~A.} \bibnamefont{Usmani}} \bibnamefont{and}
  \bibinfo{author}{\bibfnamefont{F.~C.} \bibnamefont{Khanna}},
  \bibinfo{journal}{J.Phys. G} \textbf{\bibinfo{volume}{35}},
  \bibinfo{pages}{025105} (\bibinfo{year}{2008}),
  \urlprefix\url{http://stacks.iop.org/0954-3899/35/i=2/a=025105}.

\bibitem[{\citenamefont{Ferrari~Ruffino
  et~al.}(2017)\citenamefont{Ferrari~Ruffino, Barnea, Deflorian, Leidemann,
  Lonardoni, Orlandini, and Pederiva}}]{fer2017}
\bibinfo{author}{\bibfnamefont{F.}~\bibnamefont{Ferrari~Ruffino}},
  \bibinfo{author}{\bibfnamefont{N.}~\bibnamefont{Barnea}},
  \bibinfo{author}{\bibfnamefont{S.}~\bibnamefont{Deflorian}},
  \bibinfo{author}{\bibfnamefont{W.}~\bibnamefont{Leidemann}},
  \bibinfo{author}{\bibfnamefont{D.}~\bibnamefont{Lonardoni}},
  \bibinfo{author}{\bibfnamefont{G.}~\bibnamefont{Orlandini}},
  \bibnamefont{and} \bibinfo{author}{\bibfnamefont{F.}~\bibnamefont{Pederiva}},
  \bibinfo{journal}{Few-Body Syst.} \textbf{\bibinfo{volume}{58}},
  \bibinfo{pages}{113} (\bibinfo{year}{2017}).

\bibitem[{\citenamefont{Marcucci}(2018)}]{HHsym}
\bibinfo{author}{\bibfnamefont{L.}~\bibnamefont{Marcucci}}
  (\bibinfo{year}{2018}).

\bibitem[{\citenamefont{Nogga et~al.}(2003)\citenamefont{Nogga, Kievsky,
  Kamada, Gl\"ockle, Marcucci, Rosati, and Viviani}}]{per}
\bibinfo{author}{\bibfnamefont{A.}~\bibnamefont{Nogga}},
  \bibinfo{author}{\bibfnamefont{A.}~\bibnamefont{Kievsky}},
  \bibinfo{author}{\bibfnamefont{H.}~\bibnamefont{Kamada}},
  \bibinfo{author}{\bibfnamefont{W.}~\bibnamefont{Gl\"ockle}},
  \bibinfo{author}{\bibfnamefont{L.~E.} \bibnamefont{Marcucci}},
  \bibinfo{author}{\bibfnamefont{S.}~\bibnamefont{Rosati}}, \bibnamefont{and}
  \bibinfo{author}{\bibfnamefont{M.}~\bibnamefont{Viviani}},
  \bibinfo{journal}{Phys. Rev. C} \textbf{\bibinfo{volume}{67}},
  \bibinfo{pages}{034004} (\bibinfo{year}{2003}),
  \urlprefix\url{https://link.aps.org/doi/10.1103/PhysRevC.67.034004}.

\bibitem[{\citenamefont{Friar et~al.}(1990)\citenamefont{Friar, Gibson, and
  Payne}}]{fri1990}
\bibinfo{author}{\bibfnamefont{J.~L.} \bibnamefont{Friar}},
  \bibinfo{author}{\bibfnamefont{B.~F.} \bibnamefont{Gibson}},
  \bibnamefont{and} \bibinfo{author}{\bibfnamefont{G.~L.} \bibnamefont{Payne}},
  \bibinfo{journal}{Phys. Rev. C} \textbf{\bibinfo{volume}{42}},
  \bibinfo{pages}{1211} (\bibinfo{year}{1990}).

\end{thebibliography}
\end{document}